\documentstyle[aps,twocolumn]{revtex}

\newcommand{\rv}{{\bf r}}
\newcommand{\Ev}{{\bf E}}

\newcommand{\beq}{\begin{equation}}
\newcommand{\eeq}{\end{equation}}
\newcommand{\bea}{\begin{eqnarray}}
\newcommand{\eea}{\end{eqnarray}}
\newcommand{\<}{\langle}
\renewcommand{\>}{\rangle}

\begin{document}

\draft
\preprint{}
\title{Bose-Einstein condensate in a double-well potential\\
as an open quantum system}
\author{Janne Ruostekoski and Dan F. Walls} 
\address{ Department of Physics,
University of Auckland, Private Bag 92019, 
%\\
Auckland, New Zealand}
\date{\today}
\maketitle
\begin{abstract}
We study the dynamics of a Bose-Einstein condensate in a double-well potential
in the two-mode approximation. The dissipation of energy from the condensate is
described by the coupling to a thermal reservoir of non-condensate modes. As a
consequence of the coupling the self-locked population imbalance in the
macroscopic quantum self-trapping decays away. We show that a coherent state
predicted by spontaneous symmetry breaking is not robust and decoheres
rapidly into a statistical mixture due to the interactions between condensate and
non-condensate atoms. However, via stochastic simulations we find that with a
sufficiently fast measurement rate of the relative phase between the two wells the
matter wave coherence is established even in the presence of the decoherence.

\end{abstract}

Bose-Einstein condensates (BEC's) exhibit a macroscopic quantum coherence which
in thermal atomic ensembles is absent \cite{coh}. In the conventional reasoning
the BEC is assigned a macroscopic wave function with an arbitrary but fixed phase.
The selection of a phase implies the spontaneous breaking of the gauge symmetry.
The atom-atom interactions in finite-sized BEC's affect coherence properties. The
relative phase of BEC's undergoes diffusion or collapses due to the condensate
self-interactions \cite{WRI97,LEW96}. The interactions between condensate and
non-condensate atoms create decoherence \cite{ZUR91}. Modelling decoherence by
fully including the quantum effects requires sophisticated theoretical studies
which non-trivially include non-condensate atoms. In the experiments on BEC's of
dilute alkali atomic gases \cite{becexp} trapped atoms are evaporatively cooled
and continuously exchange particles with their environment. Thus, standard
approaches of quantum optics for open systems involving master equations and heat
reservoirs seem especially natural for treating atomic BEC's
\cite{GAR97,ANG97,GAR98,JAK98,WIS98}.

In this paper we study the evolution of the master equation for a BEC in a
double-well potential in a two-mode approximation using previously derived
models \cite{ANG97}. Macroscopic quantum coherence of BEC's results in coherent
quantum tunneling of atoms between the two modes or the ``two BEC's", which is
analogous to the coherent tunneling of Cooper pairs in a Josephson junction
\cite{JAV86,DAL96,MIL97,SME97,ZAP98}. According  to the Josephson effect, the 
atom numbers of the BEC's oscillate even if the number of atoms in each
condensate is initially equal. Even BEC's with a well-defined number of atoms,
and with no phase information, could exhibit oscillations in particular
measurement processes on atoms \cite{JAC96} or on photons \cite{RUO97d,COR98}. 

One interesting feature of the coherent quantum tunneling between two BEC's
is that due to the nonlinearity arising from atom-atom interactions,
oscillations are expected to be suppressed when the population difference
exceeds a critical value in a process known as macroscopic quantum
self-trapping (MQST) \cite{MIL97,SME97}. We show that in the presence of
collisions between the condensate and non-condensate atoms MQST decays away,
i.e., the atom numbers of the two BEC's become balanced.

Anglin \cite{ANG97} derived the master equation for a trapped BEC by considering a
special model of a BEC confined in a deep but narrow spherical square-well
potential, tuned so as to posses exactly one single-particle bound state. In
this model the thermal reservoir of non-condensate atoms consists of a
continuum of unbound modes obtained by the scattering solutions of the
potential well. The binding energy $E_b$ of the BEC mode is assumed to be large
compared to the thermal energy of the noncondensed atoms. Then the fugacity
satisfies $z=e^{\beta\mu}\ll1$, where $\mu$ is the chemical potential and
$\beta=1/(k_BT)$, even in the presence of a BEC. The condensation occurs because
of the depth of the attractive potential. The small fugacity allows the
derivation of the master equation in the Markov and Born approximations. By
expanding in terms of the small parameters $z$ and $a/d$, where $a$ is the
$s$-wave scattering length and $d$ is the length scale of the BEC, the reduced
density operator for the BEC satisfies the following equation of motion:    
\bea
\dot{\rho} &=& i/\hbar [\rho,H_S]+C_1{\cal D}[b^\dagger b]\rho +C_2 {\cal D}[
b^\dagger]\rho \nonumber\\ 
&&+C_2 \exp{\{ \beta(\hbar\Delta -\mu)/2 \} }{\cal D}[b]\rho+
{\cal O}(z^2a^3/d^3)\,.
\label{eq:master}
\eea
Here we have defined
\beq
{\cal D}[c]\rho \equiv c\rho c^\dagger -1/2 (c^\dagger c\rho +
\rho c^\dagger c)\,,
\label{eq:ope}
\eeq
and $\Delta\simeq 2\kappa N-E_b$ is expressed in terms of the strength of the
self-interaction energy $\kappa$. The system Hamiltonian for the BEC is
denoted by $H_S$. For simplicity, the interactions between different
non-condensate atoms are estimated by the Boltzmann scattering rate
$\gamma=\sigma n\hbar k/m$, where $\sigma$ is the scattering cross-section and
$n$ is the density of the gas. The parameters $C_1$ and $C_2$ may then be
estimated by $C_1\sim\gamma$ and $C_2/C_1\simeq z/(\beta E_b)$.

In a single mode approximation the processes in which a BEC atom collides with a
non-condensate atom and produces two BEC particles, or vice versa, do not
conserve the energy and they are absent in Eq.~{(\ref{eq:master})}. The term
proportional to $C_1$ describes elastic two-body collisions between condensate and
non-condensate atoms and it induces phase decoherence \cite{WAL85}. Inelastic
collisions [the terms proportional to $C_2$ in Eq.~{(\ref{eq:master})}] introduce
amplitude decoherence \cite{WAL85}. With the present approximations the amplitude
decoherence is dramatically reduced compared to the phase damping. The central
assumption is a small fugacity indicating $\beta E_b\gg 1$. In the scattering
processes the energy must be conserved to leading order, so that depletion and
growth of the BEC involve non-condensate atoms with high enough energy to balance
the large binding energy of the trapped state. If the BEC is described by a
multiple number of modes \cite{GAR98}, the amplitude decoherence does not need to
be small compared to the phase decoherence. The necessary condition for the
validity of the single-mode approximation in a harmonic trap is that the
self-energy of the atom-atom interactions of the BEC should not dominate over the
mode energy spacing. This assumption clearly breaks down in the Thomas-Fermi limit
in which the kinetic energy is negligible compared to the self-interaction energy
indicating $(15Na/l)^{1/5}\gg 1$, where $l=(\hbar/m\omega)^{1/2}$ is the length
scale of the harmonic oscillator. Jaksch {\it et al.} \cite{JAK98} have
calculated the intensity and the amplitude fluctuations of a BEC. They have
evaluated the coefficients of the master equation in the Thomas-Fermi limit and
have obtained stronger amplitude decoherence than phase decoherence.
Nevertheless, the calculations of the BEC fluctuations are still performed in the
one-mode approximation.

We consider Anglin's model for the master equation \cite{ANG97} in the studies
of the coherent quantum tunneling of a BEC in a double-well potential. To obtain
the system Hamiltonian for the BEC we approximate the total field operator by the
two lowest quantum modes and write $\psi(\rv)\simeq \phi_b(\rv) b +\phi_c(\rv)
c$, where $\phi_b$ and $\phi_c$ are the local mode solutions of the individual
wells with small spatial overlap and the corresponding annihilation operators are
$b$ and $c$. This requires that the BEC self-interaction energy should not
dominate over the energy separation of these modes from the higher excited modes.
The Hamiltonian in the two-mode approximation reads \cite{MIL97}       
\beq
H_S/\hbar=\xi b^\dagger b+\Omega (b^\dagger c + c^\dagger b)+
\kappa [ (b^\dagger)^2 b^2 + (c^\dagger)^2 c^2]\,.
\label{eq:ham}
\eeq
Here $\xi$ is the energy difference between the modes. The tunneling
between the two wells is described by $\Omega$, which is proportional to the
overlap of the spatial mode function of the opposite wells. The short-ranged
two-body interaction strength is obtained from $\kappa=2\pi a\hbar /m \int
|\phi_b (\rv)|^4$, where we have assumed $\int |\phi_b (\rv)|^4= \int |\phi_c
(\rv)|^4$.

It is useful to describe the dynamics of Eq.~{(\ref{eq:ham})} in terms of the
atomic coherent states \cite{ARE72}, which exhibit coherence, but conserve the
total number of atoms:
\beq 
|\theta,\phi\rangle=\sum_{m=-j}^j {2j\choose
m+j}{\tau^{m+j}\over (1+|\tau|^2)^j}\, |j,m\rangle \,, 
\label{eq:atomcoh}
\eeq 
where $|j,m\rangle $ is an eigenstate of the angular momentum operators
$\hat{J}_z$ and $\hat{J}^2$ with eigenvalues $m$ and $j(j+1)$. Here $\tau\equiv
e^{-i\phi}\tan{(\theta/2)}$. We define the angular momentum operators in terms of
the BEC operators in the usual way: $\hat{J}_x =(b^\dagger {c}+c^\dagger b)/2 $,
$\hat{J}_y = (  b^\dagger {c}-c^\dagger b)/(2i)$, and $\hat{J}_z =  (  b^\dagger
{b}-c^\dagger c)/2$. Then $j=N/2$ is a constant of motion and the atomic coherent
states may be expressed in terms of the number states of the BEC's 
\beq
|\theta,\phi\rangle_N= \sqrt{{N!\over N^N}}\sum_{l=0}^N
{{\cal C}^{N-l}{\cal B}^l\over\sqrt{l!(N-l)!}} |l,N-l\rangle\,,
\label{eq:proj}
\eeq
where $l=m+N/2$, and ${\cal B}$ and ${\cal C}$ are the `coherent amplitudes' of
the two BEC's: $\langle b^\dagger b \rangle = |{\cal B}|^2$, $\langle c^\dagger
c\rangle = |{\cal C}|^2$, and  $\langle b^\dagger c\rangle = |{\cal B}| |{\cal
C}| e^{i\varphi}$. The relative phase between the two wells is $\varphi$. Equation
(\ref{eq:proj}) clearly shows how the atomic coherent states  are
projections of the coherent states $|{\cal B},{\cal C}\rangle$ onto the basis sets
of a fixed total number of atoms $N$.

We study the evolution of atomic coherent states in the presence of decoherence
in both wells $b$ and $c$. As already noted, with the present
approximations the amplitude decoherence is much weaker than the phase
decoherence. It is also easy to verify numerically that the coherent states are
much less robust for the phase decoherence than for the amplitude decoherence,
even if the magnitude of the phase and the amplitude damping is equal. This is
also well known in quantum optics \cite{WAL85}. Hence, the effects of the
amplitude decoherence are negligible compared to the phase decoherence.

The dynamics of the master equation is studied in terms of stochastic
trajectories of state vectors \cite{DAL92}. The master equation is unraveled
by Monte-Carlo evolution of wave functions.

In Fig. \ref{fig:1}a) we have plotted the expectation value of the number of
atoms in the well $b$, $N_b(t)$. The initial state is the atomic coherent state
with the relative phase $\varphi=\pi/2$ between the two wells and the
expectation values for the atom numbers $N_b=N_c=50$. We have
set $N\kappa/\Omega = 0.5$ and $\xi=0$. The solid line is the result without
decoherence $\gamma=0$. The oscillations are damped due to the collapse of the
macroscopic coherence \cite{MIL97}. The dashed line has  $\gamma=0.1\kappa$ and
the dashed-dotted line $\gamma=0.4\kappa$. The decoherence clearly increases the
damping of the oscillations. Although the model used is very simplified, we can
make rough estimates for the parameters. For the effective mode volume $1/\int
|\phi(\rv) |^4=10^{-9}$ cm$^3$, $a=5$ nm, for $^{23}$Na, and for the temperature
$T=100$ nK, $\gamma=0.1\kappa$ corresponds to the density of the non-condensate
atoms $n\sim 10^{9}$ cm$^{-3}$ and the fugacity $z\sim10^{-3}$. In Fig.
\ref{fig:1}b) we have plotted ${\rm Tr}(\rho^2)$ for the same run of simulations.
We see that a pure state predicted by spontaneous symmetry breaking is not robust
and decoheres rapidly into a statistical mixture due to the interactions between
condensate and non-condensate atoms. 

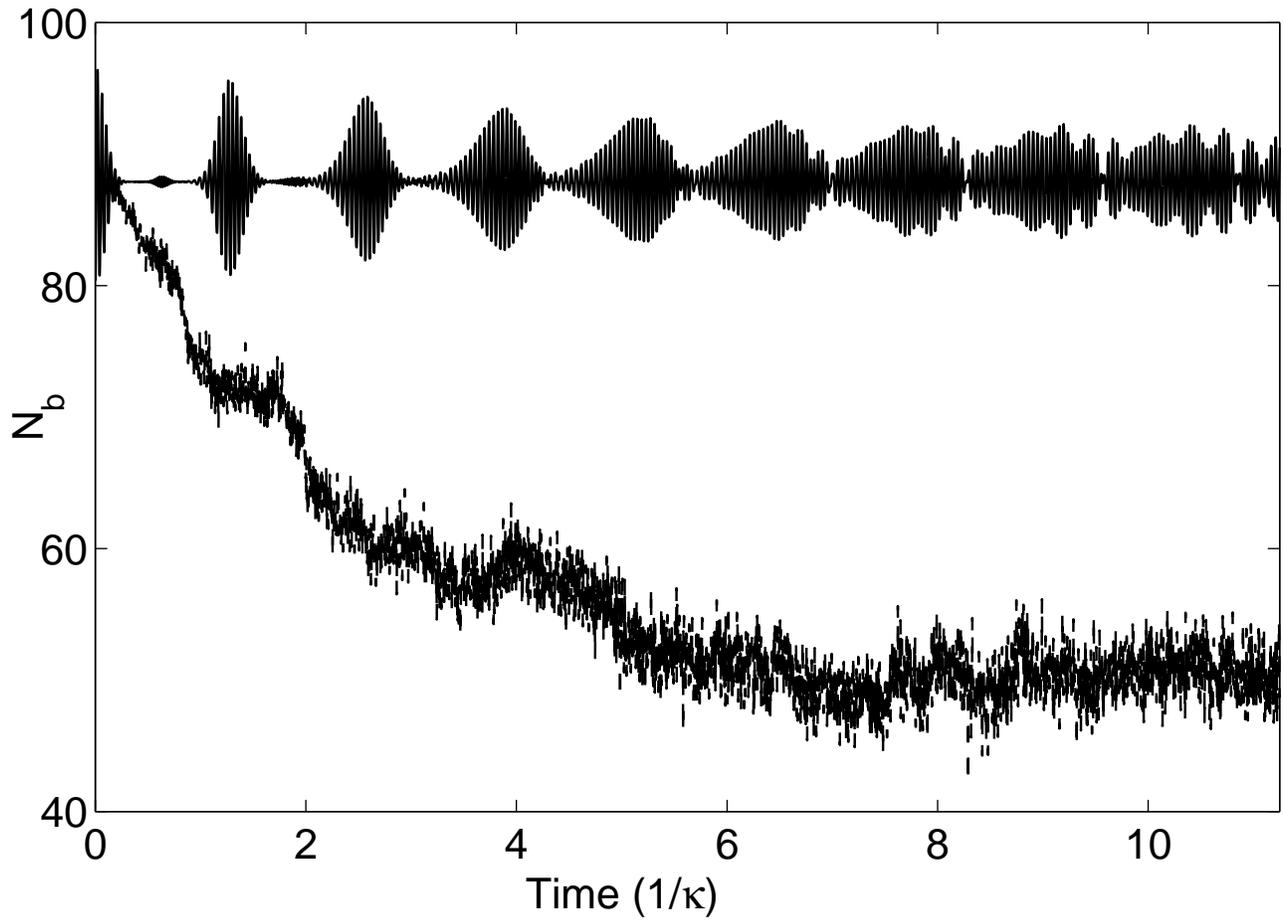
\begin{figure}
\caption{
The expectation value of the number of atoms in the well $b$ in the case of large
nonlinearity. Initial state is the atomic coherent state with $N_b=4N_c=80$ and
$\varphi=0$. The solid line is the result without decoherence and describes the
macroscopic quantum self-trapping. The oscillations undergo collapses and
revivals. The dashed line shows how the atom population becomes balanced in the
presence of decoherence.
 }
\label{fig:2}
\end{figure}

If the nonlinearity is large compared to the tunneling frequency and the
population imbalance exceeds a critical value, the oscillations of the atom
numbers are suppressed \cite{MIL97,SME97}. A large number of atoms remains
``locked" in one of the wells. In Fig. \ref{fig:2}a) we have plotted $N_b(t)$
obtained by integrating Eq.~{(\ref{eq:ham})} (the solid line) and the solution of
the master equation in the presence of the decoherence $\gamma=0.2\kappa$ (the
dashed line). In this case $N\kappa/\Omega = 4.5$, $\xi=0.005\Omega$, and the
initial state is the atomic coherent state with the expectation values
$N_b=4N_c=80$ and $\varphi=0$. Due to the interactions between condensate and
non-condensate atoms MQST vanishes and the atom population becomes balanced.

Next, we include the effect of measurements into the calculations. We assume
that the number of atoms is nondestructively measured in one of the two wells.
The effect of the measurement is included in quantum trajectory simulations by
averaging over the dissipation channels corresponding to the interactions between
condensate and non-condensate atoms, but at the same time by considering the
measurement of the number of atoms in one of the wells to be a single realization
of a stochastic trajectory. We consider a particular situation in which the
Josephson dynamics is nondestructively measured by shining a coherent light beam
through one of the BEC's. 

We assume that the incoming light field with a large detuning from the atomic
resonance is scattered from the well $b$. For instance, if the shape of the gas is
flat and the light is shined through a thin dimension, the dipole shifts are
small and the sample can be considered optically thin \cite{JAV95b}. A BEC atom
scatters back to the BEC via coherent spontaneous scattering, stimulated by a
large number of atoms in the BEC. Coherently scattered photons are emitted into a
narrow cone in the forward direction. The  amplitude of the scattered field has
the dependence $|\Ev^+_S|\propto {\cal E} d_{eg}/(\hbar\Delta) \, b^\dagger b$ on
the detuning $\Delta$, the dipole matrix element $d_{eg}$, and the amplitude of
the incoming field $ {\cal E}$ \cite{JAV95b}. The direct counting of
spontaneously emitted photons can be simulated in terms of quantum trajectories
\cite{DAL92}, in which the stochastic quantum ``jumps" correspond to the
detections of photons. The procedure is similar to Ref. \cite{RUO98a}. The
detection rate of the scattered photons in the present case is $ \Gamma \< 
(b^\dagger b)^2 \> \propto |\Ev^-_S\cdot \Ev^+_S|$.

If the number of atoms in a BEC is not large, the scattering between the
condensate and non-condensate modes is not negligible. This introduces
amplitude decoherence similar to the amplitude decoherence due to the atomic
collisions in Eq.~{(\ref{eq:master})}. If we require that the two-mode
approximation describes accurately the tunneling dynamics, for small harmonic
traps the amplitude decoherence due to the light scattering may not be
negligible. For large traps the two-mode approximation can be accurate even for
large atom numbers as the BEC self-interaction energy $\kappa \propto l^{-3}$ and
the trap frequency $\omega\propto l^{-2}$. Nevertheless, as a first approximation
we ignore the decoherence due to the light scattering.

The density matrix of a BEC may be reconstructed by the nondestructive
measurements of the number of atoms in one of the wells. The procedure is
similar to Ref. \cite{BOL97}, except that the density matrix has now time
dynamics determined by the Hamiltonian (\ref{eq:ham}), the dissipation, and the
back-action of the measurements. Following the notation of Ref. \cite{BOL97}, at
the time $t$ we have $\rho(t)=\hat{U}(t)^\dagger \rho \hat{U}(t)$, where 
$\hat{U}(t)$ is in this case the time propagator. Then the probability of the
measurement result of $m$ atoms in the well $b$ at the time $t$ is given by
$P_m(t)=\< m| \rho(t)|m \>$. By inverting this expression the density matrix can
be reconstructed \cite{BOL97}.

The off-diagonal long range order (ODLRO) between the two wells may be described
by the visibility of the interference $\beta$ \cite{RUO97d}. To emphasize the
effect of decoherence on $\beta$ we ignore the oscillating dynamics of $H_S$
(including the collapses and revivals) by propagating back the system dynamics. In
accordance with Ref. \cite{RUO97d} we define 
\beq
\beta e^{i\varphi}\equiv {2\over N}{\rm Tr}[e^{ iH_S t/\hbar }\rho
e^{ -iH_S t/\hbar } b^\dagger (0) c(0) 
] \,.
\eeq 
For a coherent state we have $\beta=1$, and $\varphi$ is the relative phase
between the two wells. However, for a number state there is no phase information
and $\beta=0$. If the BEC's have unequal atom numbers, the maximum visibility is
reduced from one to $\beta_{\rm max}=2\sqrt{N_bN_c}/N$. Hence, it is useful to
define the relative visibility by $\beta_r\equiv\beta /\beta_{\rm max}$.

We simulate the dynamics of the dissipation and the measurements by repeating
single realizations of quantum trajectories. In the first realization we save
the stochastic times of the photon detections. In every subsequent run of the
trajectory the times of the photon detections are forced to be the same as in
the first run. Although the photon detection times after the first trajectory
are deterministic, the collision times between condensate and non-condensate atoms
corresponding to the dissipation channels are stochastic in every run.
Averaging over all the trajectories allows us to consider the photon
measurements to be a ``single realization" of the quantum trajectory even though
the atomic collisions are at the same time ensemble averages corresponding to the
density matrix evolution.

We consider a situation that the two BEC's are initially in pure number states
with $N_b=52$ and $N_c=48$. We have set $N\kappa/\Omega = 0.25$,
$\xi=0.005\Omega$, and the photon scattering rate $\Gamma=0.8\kappa$. In Fig.
\ref{fig:3}a) we have plotted $\beta_r(t)$. The solid line is the result without
decoherence $\gamma=0$. The dashed line has  $\gamma=0.05\kappa$ and the
dashed-dotted line $\gamma=1.8\kappa$. In the beginning for the number state
$\beta_r=0$, but $\beta_r \rightarrow 1$ rapidly even in the presence of weak
decoherence. As a consequence of the decoherence  ODLRO starts decreasing, but
the measurements of spontaneously scattered photons establish the macroscopic
coherence, even though the BEC's are initially in pure number states. In Fig.
\ref{fig:3}b) we have plotted  ${\rm Tr}(\rho^2)$ for the same run. We see that
${\rm Tr}(\rho^2)$ remains close to one and the state is reasonably pure due to
the fast measurement rate even in the presence of decoherence if
$\gamma=0.0625\Gamma$. In the case of stronger decoherence with
$\gamma=2.25 \Gamma$ the state evolves into a statistical mixture.

In conclusion, we have shown that as a consequence of the interactions between
condensate and non-condensate atoms MQST decays away. Due to the interactions a
BEC does not remain in a pure state with a well-defined relative phase. However,
the coherence properties can be established via the measurement process even in
the presence of decoherence. In particular, nondestructive detections allow
the measurements of phase dynamics.

We acknowledge discussions with A. C. Doherty. This work was supported by the
Marsden Fund of the Royal Society of NZ and The University of Auckland Research
Fund.

\begin{figure}
\caption{
The expectation value of the number of atoms in the well $b$ in a) as a function
of time. Initial state is the atomic coherent state with $N_b=N_c=50$ and the
relative phase $\varphi=\pi/2$. The solid line is the result without decoherence
$\gamma=0$. The dashed line has  $\gamma=0.1\kappa$ and the dashed-dotted line
$\gamma=0.4\kappa$. In b) ${\rm Tr}(\rho^2)$ for the same run. }   
\label{fig:1}   
\end{figure}

\begin{figure}
\caption{
The relative visibility of the interference $\beta_r(t)$ in a) when the number of
atoms in one well is nondestructively measured by light scattering. The BEC's are
initially in pure number states with $N_b=52$ and $N_c=48$. The photon
scattering rate $\Gamma=0.8\kappa$. The solid line is the result without
decoherence $\gamma=0$. The dashed line has $\gamma=0.05\kappa$ and the
dashed-dotted line $\gamma=1.8\kappa$. In b) ${\rm Tr}(\rho^2)$ for the same run.
 }
\label{fig:3}
\end{figure}


\begin{references}

\bibitem{coh} M. R. Andrews {\it et al.}, Science {\bf 275}, 637 (1997); E. A.
Burt {\it et al.}, Phys. Rev. Lett. {\bf 79}, 337 (1997).

\bibitem{WRI97} E. M. Wright {\it et al.}, Phys. Rev. A {\bf 56}, 591 (1997).

\bibitem{LEW96} M. Lewenstein and L. You, Phys. Rev. Lett. {\bf 77}, 3489
(1996); Y. Castin and J. Dalibard, Phys. Rev. A {\bf 55}, 4330
(1997); J. Javanainen and M. Wilkens, Phys. Rev. Lett. {\bf 78}, 4675
(1997).

\bibitem{ZUR91} W. H. Zurek, Phys. Today {\bf 44} (10), 36 (1991) and references
therein.

\bibitem{becexp} M. H. Anderson {\it et al.},  Science {\bf 269}, 198
(1995); K. B. Davis {\it et al.}, Phys. Rev. Lett. {\bf  75}, 3969 (1995);
C. C. Bradley {\it et al.}, Phys. Rev. Lett. {\bf 78}, 985 (1997).

\bibitem{GAR97} D. Jaksch, C. W. Gardiner, and P. Zoller, Phys. Rev. A {\bf 56},
575 (1997).  

\bibitem{ANG97} J. Anglin, Phys. Rev. Lett. 79, 6 (1997).

\bibitem{GAR98} C. W. Gardiner and P. Zoller, cond-mat/9712002.

\bibitem{JAK98} D. Jaksch {\it et al.}, cond-mat/9712206. 

\bibitem{WIS98} H. M. Wiseman and J. A. Vacaro, unpublished.

\bibitem{JAV86} J. Javanainen, Phys. Rev. Lett. {\bf 57}, 3164
(1986). 

\bibitem{DAL96} F. Dalfovo, L. Pitaevskii, and S. Stringari, Phys. Rev. A {\bf
54}, 4213 (1996).  

\bibitem{MIL97} G. J. Milburn {\it et al.}, Phys. Rev. A {\bf 55}, 4318 (1997).

\bibitem{SME97} A. Smerzi {\it et al.}, Phys. Rev. Lett. {\bf 79}, 4950 (1997).

\bibitem{ZAP98} I. Zapata, F. Sols, and A. J. Leggett, Phys. Rev. A {\bf 57} R28
(1998). 

\bibitem{JAC96} M. W. Jack, M. J. Collett, and D. F. Walls, Phys. Rev. A
{\bf 54}, R4625 (1996). 

\bibitem{RUO97d} J. Ruostekoski and D. F. Walls, Phys. Rev. A {\bf 56}, 2996
(1997). 

\bibitem{COR98} J. F. Corney and G. J. Milburn, cond-mat/9712282.

\bibitem{WAL85} D. F. Walls and G. J. Milburn, Phys. Rev. A {\bf 31},
2403 (1985).

\bibitem{ARE72} F. T. Arecchi {\it et al.}, Phys. Rev. A {\bf 6}, 2211 (1972). 

\bibitem{DAL92} J. Dalibard, Y. Castin, and K. M\o lmer, J. Opt. Soc. Am. {\bf
10}, 524 (1993) and references therein.

\bibitem{JAV95b} J. Javanainen and J. Ruostekoski, Phys. Rev. A {\bf 52}, 3033
(1995).

\bibitem{RUO98a} J. Ruostekoski {\it et al.},  Phys. Rev. A {\bf 57}, 511
(1998).

\bibitem{BOL97} E. L. Bolda, S. M. Tan, and D. F. Walls, Phys. Rev. Lett. {\bf
79}, 4719 (1997); R. Walser, {\it ibid.}, 4724 (1997).


\end{references}
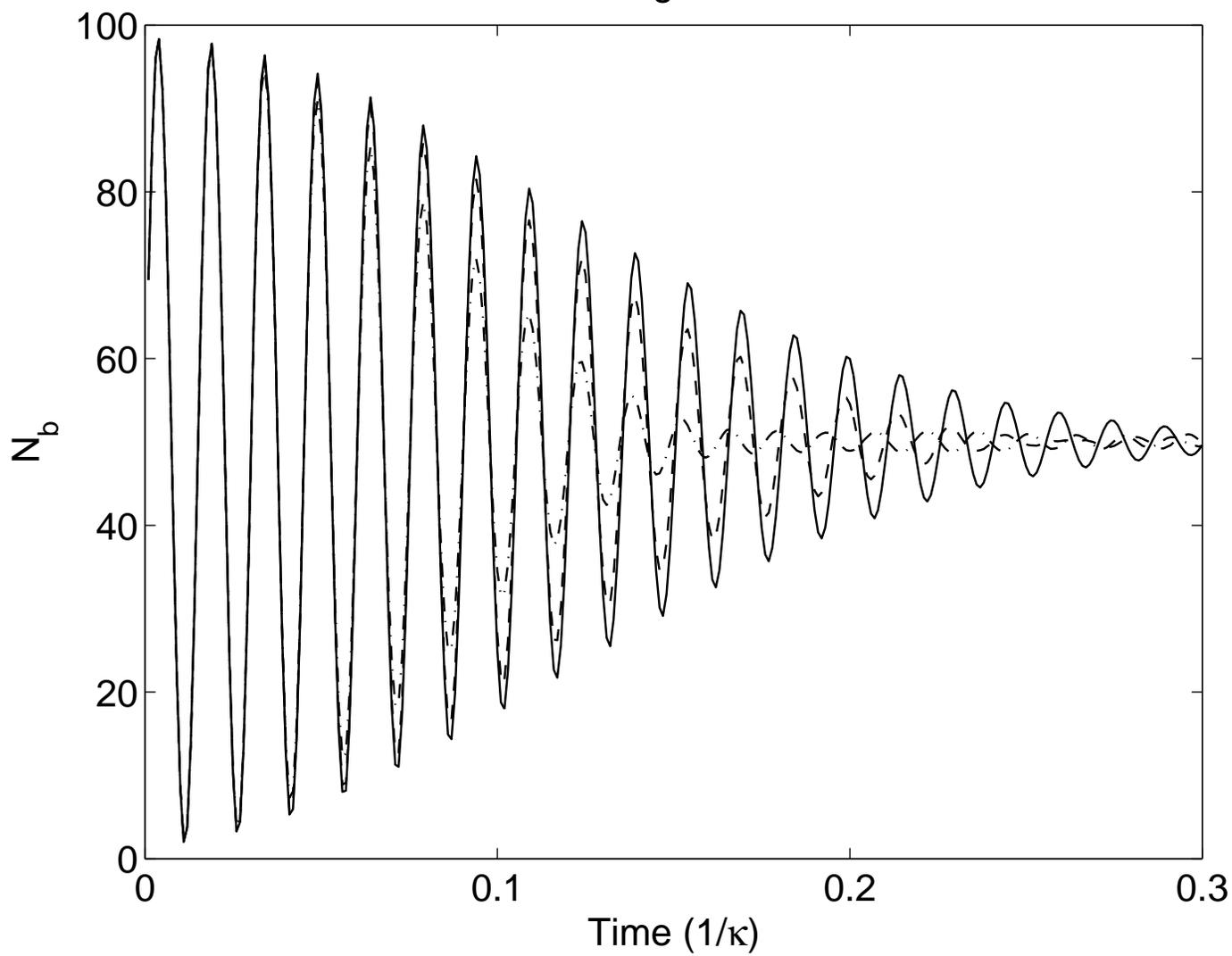
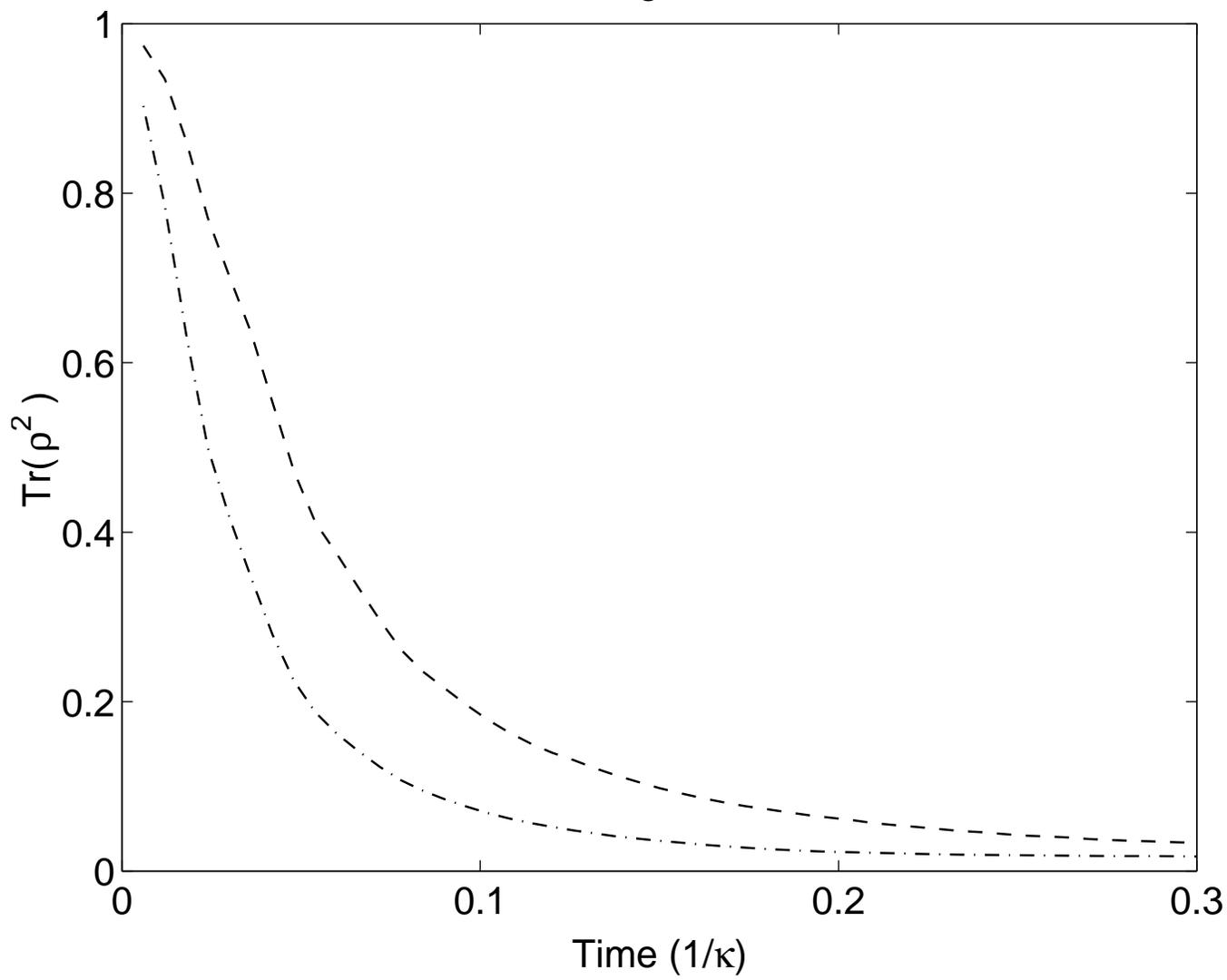
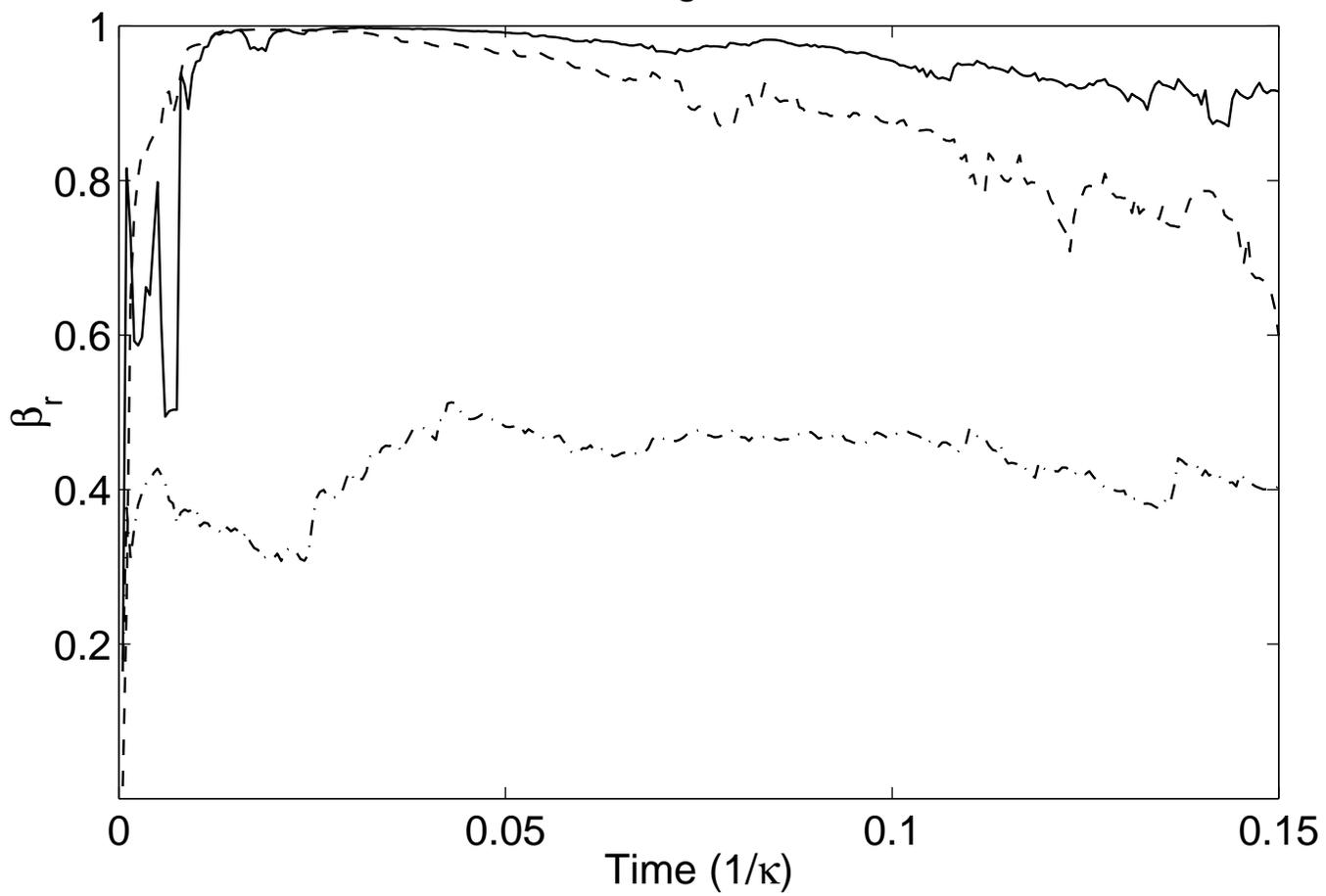
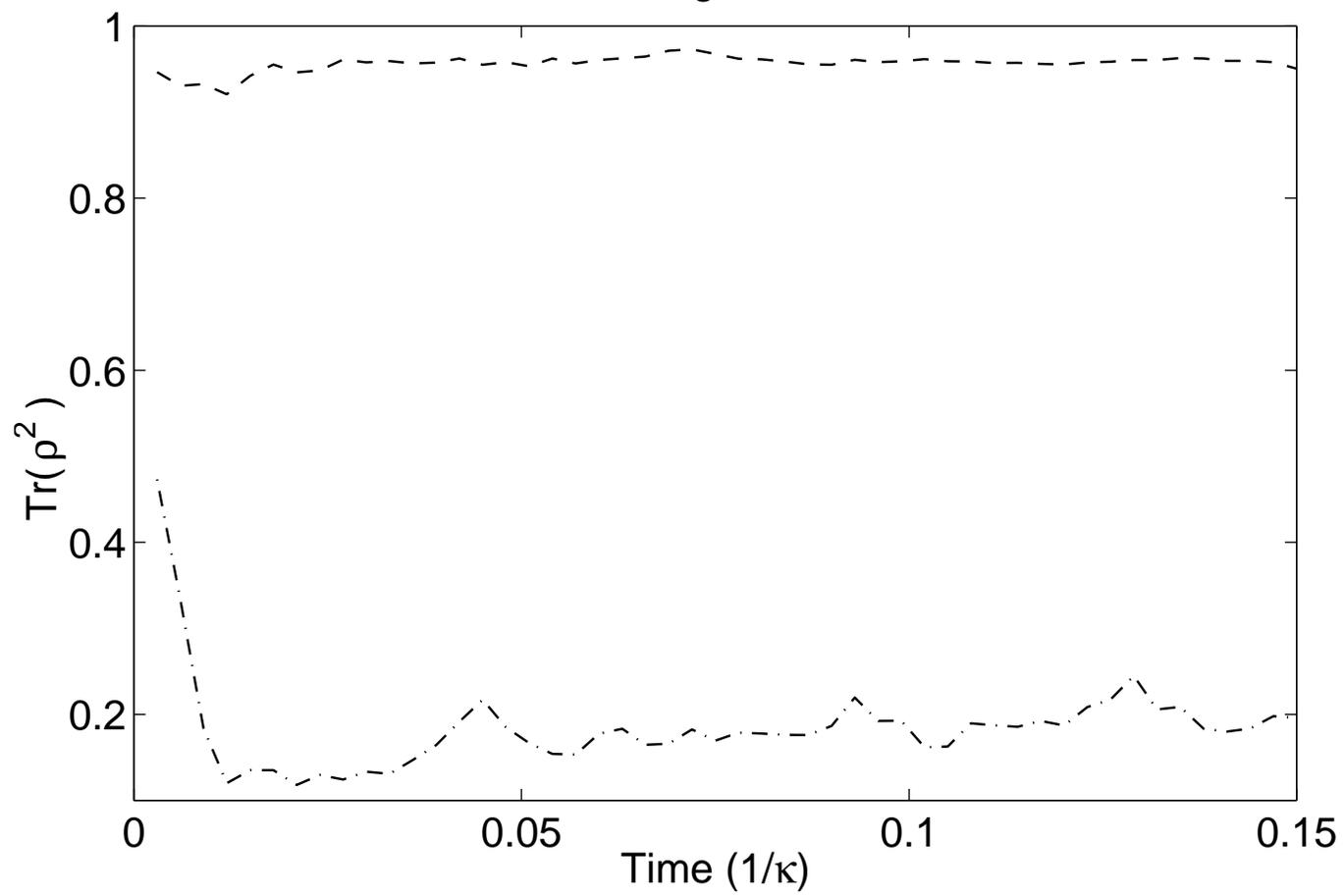
\end{document}